\documentclass[prmaterials,twocolumn,aps,reprint,floatfix,superscriptaddress]{revtex4-2}

\pdfoutput=1
\usepackage{graphicx,color}
\usepackage[center]{subfigure}
\usepackage{mathtools}
\usepackage{bm}
\usepackage{esvect}
\usepackage{enumitem}
\usepackage{wrapfig}
\usepackage{ulem}
\usepackage[mathscr]{euscript}

\newcommand \gbound {domain wall\ }
\newcommand \jeng {\chi}
\newcommand \ff {F}

\newcommand \be {\begin{eqnarray}}
\newcommand \ee {\end{eqnarray}}
\newcommand \ben {\begin{eqnarray}}
\newcommand \een {\end{eqnarray}}
\newcommand \Lap {\nabla^2}

\newcommand \nline {\nonumber \\}

\newcommand \etal {{\it et al.\ }}
\newcommand \pt[2] {\frac{\partial #1}{\partial #2}}

\begin{document}

\title{Moir\'e patterns and inversion boundaries in graphene/hexagonal boron nitride bilayers}

\author{K. R. Elder}
\affiliation{Department of Physics, Oakland University, Rochester, Michigan 48309, USA}

\author{Zhi-Feng Huang}
\affiliation{Department of Physics and Astronomy, Wayne State University, Detroit, Michigan 48201, USA}

\author{T. Ala-Nissila}
\affiliation{QTF Centre of Excellence, Department of Applied Physics, 
Aalto University School of Science, P.O. Box 11000, FI-00076 Aalto, Espoo, Finland}
\affiliation{Department of Physics, P.O. Box 1843, Brown University,
Providence, RI 02912-1843, USA}
\affiliation{Interdisciplinary Centre for Mathematical Modelling, 
Department of Mathematical Sciences, Loughborough University, 
Loughborough, Leicestershire LE11 3TU, UK}

\begin{abstract}
In this paper a systematic examination of graphene/hexagonal boron nitride (g/hBN) 
bilayers is presented, through a recently developed two-dimensional phase field 
crystal model that incorporates out-of-plane deformations. The system parameters 
are determined by closely matching the stacking energies and heights of 
graphene/hBN bilayers to those obtained from existing quantum-mechanical density functional 
theory calculations. Out-of-plane deformations are shown to reduce the energies
of inversion domain boundaries in hBN, and the coupling between graphene and hBN
layers leads to a bilayer defect configuration consisting of an inversion 
boundary in hBN and a \gbound in graphene. Simulations of twisted 
bilayers reveal the structure, energy, and elastic properties of the
corresponding Moir\'e patterns, and show a crossover, as the misorientation 
angle between the layers increases, from a well-defined hexagonal network 
of domain boundaries and junctions to smeared-out patterns.  The transition 
occurs when the thickness of domain walls approaches the size of the Moir\'e 
patterns, and coincides with the peaks in the average von Mises and volumetric 
stresses of the bilayer.
\end{abstract}

\maketitle

\section{Introduction}

Two-dimensional (2D) materials, such as graphene (g), hexagonal boron nitride (hBN), 
and transition metal dichalcogenides (TMDs), have been of continuously great 
interest in recent years due to their extraordinary electronic, thermal, and 
mechanical properties and potential for various technological applications 
\cite{AkinwandeNature19,Ajayan16}. Currently there has been a focus on 
stacking of such materials together to form multiple-layer structures with tunable physical properties.  
Perhaps the simplest of such systems, namely a graphene bilayer, has long shown 
interesting behavior ranging from being a good insulator to a superconductor
\cite{Guinea13174,Yankowitz19}, while the stacking of an hBN layer onto 
a graphene monolayer significantly increases thermal conductivity \cite{Momeni2020}. 
Many other exotic features, particularly those arising from the modulation
of novel electronic properties, such as fractal quantum Hall effects in 
g/hBN bilayers \cite{Dean13,HuntScience13,Moon2014,Iwasaki2020}, 
have also been reported.

An important feature of bilayer heterostructures is the
emergence of Moir\'e patterns or superlattices which play a key role in 
determining the material properties described above, given their long-range
superstructural behavior of periodic structural and electronic modulations coupled with the underlying 
short-range atomic-scale lattice or sublattice structure 
\cite{Dean13,HuntScience13}. Moir\'e patterns in g/hBN bilayers 
with different twist angles have been observed in experiments 
\cite{Dean13,HuntScience13,Tang2013,Iwasaki2020,Huang2021} 
and examined in theoretical studies \cite{Moon2014,HuPRM17,Chandra2020}.
However, most of existing work has focused on a relatively narrow range of 
small misorientation angles between the two layers, while knowledge of
higher-angle Moir\'e patterns and also the elastic behavior of the bilayers
is still sparse, which limits understanding and further development of
this type of heterostructural system.  This would then require a systematic 
study of the structural, energetic, and elastic properties of the g/hBN 
bilayers across a much wider range of interlayer twist angles, as will be 
explored in this paper through efficient multi-scale modeling and simulations.

To this end, phase field crystal (PFC) models that were developed 
and parameterized for the study of 2D layers of graphene \cite{Pete16}
and hBN \cite{Taha17} will be exploited.  
Several different types of PFC models for 
graphene were examined in Ref.~\cite{Pete16} and compared with quantum-mechanical
density functional theory (DFT) and molecular dynamics (MD) calculations 
in terms of energies of grain boundaries, polycrystals, and triple 
junctions \cite{Pete16,Pete17}.  The model termed PFC1 in that work 
will be used here.  In Refs.~\cite{Taha17,Taha19} a binary PFC model 
with sublattice ordering was developed, and was used to examine 
various types of grain boundaries and defect core structures in hBN
monolayers, with results shown to be in good agreement with experiments 
and other theoretical studies.  These PFC models have been applied and
extended to study various other structural and dynamical properties of
2D materials, such as grain rotation and coupled motion in graphene and
hBN \cite{Waters22}, g/hBN lateral heterostructures 
\cite{HirvonenPRB19}, and ternary 2D hexagonal materials and in-plane 
TMD/TMD heterostructures and multijunctions \cite{Huang22}.
However, these models were strictly 2D and did not allow for out-of-plane 
variations.  Recently a simple extension of these models was 
developed to account for small out-of-plane deformations \cite{Elder21}.   
Such deformations were shown to significantly lower the energy of 
dislocations, consistent with other atomistic studies using DFT and MD.  
In addition, graphene/graphene, graphene/hBN, and hBN/hBN bilayers 
were also considered there, with the coupling between the layers 
parameterized by fitting to quantum DFT results of stacking energies 
and heights obtained by Zhou \etal \cite{Zhou15} with the use of an
analytical one-mode approximation for the PFC bilayer models. 

In this paper the previous model developed and a more accurate parameter 
fit to DFT calculations for g/hBN bilayers is used to study 
inversion domain boundaries of hBN as well as Moir\'e patterns that 
emerge when the graphene and hBN layers are rotated with respect to 
each other.  The numerical results are not only consistent with
previous experimental and theoretical findings, but also 
provide predictions for the energy density of an inversion boundary 
in graphene/hBN bilayers and 
in rotated layers for the twist-angle dependence and a transition of Moir\'e 
pattern properties and the bilayer elastic state.  

In the next 
sections a description of the model (in Sec.~\ref{sec:model}) and the 
parameterization through fitting to DFT calculations for the equilibrium 
states (Sec.~\ref{sec:equi}) are presented.  This is followed by an 
examination of inversion domain boundaries in hBN (Sec.~\ref{sec:IDB}),
showing a reduction of grain boundary energies by 8\% to 13.9\% as caused
by out-of-plane deformations, and a predicted defect configuration of 
g/hBN bilayer with an inversion boundary in the hBN layer 
coupled to a domain wall in the graphene layer. The properties of
Moir\'e patterns in twisted g/hBN bilayers are studied in
Sec.~\ref{sec:bilayers} as a function of misorientation angle, 
including the distributions of layer height difference, free energy 
density, and volumetric and von Mises stresses. Of particular focus 
is the variation in various features of the Moir\'e patterns (e.g.,
the buckling, energy profile, site occupancy, and stresses) with 
the bilayer twist angle, revealing a predicted transition to high-angle
properties (including smeared-out patterns and stress distribution)
that were unknown before. Finally, our conclusions of the
results and summary are given in Sec.~\ref{sec:summary}.

\section{Model}
\label{sec:model}

In the PFC model the free energy functional $\ff$ for a graphene/hBN bilayer can be written as \cite{Elder21}
\be
\ff = c_{\rm g} (\ff_{\rm g}+\ff_{\rm gh}) + c_{\rm h} \ff_{\rm h},
\label{eq:freet}
\ee
where $c_{\rm g}=6.58$ eV and $c_{\rm h}=2.74$ eV set the energy scales for graphene \cite{Pete16} and hBN \cite{Taha17} layers, respectively.
$\ff_{\rm g}$ is the dimensionless free energy functional for a 
flexible graphene layer, i.e.,
\be
\ff_{\rm g}&=& 
\int d\vec{r} \left[
\frac{\Delta B}{2} n_{\rm g}^2+\frac{B^{\rm x}}{2} 
\left((\nabla^2+q_{\rm g}^2) n_{\rm g}\right)^2
+\frac{\tau}{3}n_{\rm g}^3
+ \frac{v}{4}n_{\rm g}^4 \right.\nline && \left.+
\frac{\kappa}{2}  \int d\vec{r}^{\,\prime} 
C_{\rm g}(|\vec{r}-\vec{r}^{\,\prime}|) h_{\rm g}(\vec{r}) h_{\rm g}(\vec{r}^{\,\prime})\right],
\label{eq:fgraph}
\ee
where the Fourier component of $C_{\rm g}$ is given by
\be
\hat{C}_{\rm g}(k) = \left\{
\begin{array}{cc}
k^4, & k<k_{\rm max}; \\
C_{\rm max}, & k>k_{\rm max}.
\end{array}
\right.
\ee
In the limit of $\kappa=0$, Eq.~(\ref{eq:fgraph}) is the model 
termed PFC1 in Ref.~\cite{Pete16}.
In Eq.~(\ref{eq:fgraph}), $n_{\rm g}$ is \textcolor{black}{proportional} to the 
atomic number density difference that enters classical density 
functional theory in the appropriate limit \cite{Elder07}
and $h_{\rm g}$ is the height of the graphene sheet.  
The parameters entering Eq.~(\ref{eq:fgraph}) were fit 
to graphene in Refs.~\cite{Elder21,Pete16,Pete17} 
and are $\Delta B=-0.15$, $q_{\rm g}=1$,
$B^{\rm x} = v=1$, $\tau=0.8748$, $\kappa=0.114$, and the average density $\bar{n}_{\rm g}=0$. 
$\ff_{\rm h}$ is the dimensionless free energy functional of the hBN layer, 
given by \cite{Taha17,Taha19}
\be
\ff_{\rm h} &=& \int d\vec{r} 
\Big[
-\frac{\varepsilon_{N}}{2} n_{\rm N}^2+\frac{1}{2}
\left((\nabla^2+q_{\rm N}^2)n_{\rm N}\right)^2
-\frac{g_{\rm N}}{3}n_{\rm N}^3 \nline && + \frac{1}{4} n_{\rm N}^4 
-\frac{\varepsilon_{\rm B}}{2} n_{\rm B}^2+\frac{\beta_{\rm B}}{2}
\left((\nabla^2+q_{\rm B}^2)n_{\rm B}\right)^2 -\frac{g_{\rm B}}{3}n_{\rm B}^3
\nline &&  + \frac{v}{4} n_{\rm B}^4 
+\alpha_{\rm NB}n_{\rm N}n_{\rm B}+\frac{w}{2} n_{\rm N}^2 n_{\rm B} 
+ \frac{u}{2}n_{\rm N} n_{\rm B}^2 
\nline &&
+\frac{\kappa_{\rm h}}{2}\int d\vec{r}^\prime 
C_{\rm h}(|\vec{r}-\vec{r}^\prime|)h_{\rm h}(\vec{r})
h_{\rm h}(\vec{r}^\prime)
\Big],
\label{eq:fhBN}
\ee
where $n_{\rm N}$ and $n_{\rm B}$ are proportional to the atomic number density 
differences of the N and B species \cite{Taha19}, respectively, and $h_{\rm h}$ is the height of 
the hBN layer.   
The parameters have been fitted to hBN \cite{Taha17}, with
$\varepsilon_{\rm N}=\varepsilon_{\rm B} = 0.3$, 
$\alpha_{\rm NB}=0.5$, $g_{\rm N}=g_{\rm B}=0.5$, $w=u=0.3$, 
$\beta_{\rm B}=v=1$, and the average densities
$\bar{n}_{\rm N}=\bar{n}_{\rm B}=-0.28$. The bending energy coefficient $\kappa_{\rm h}$
was calculated by Guo \etal \cite{Guo2016} to be $0.89$ eV, which 
in dimensionless units corresponds to $0.32$ here. 
The values of wave numbers $q_{\rm N}$ and $q_{\rm B}$ are set to 
a common value $q_{\rm h}$ 
and will be determined in the next section. 
$\ff_{\rm gh}$ is the dimensionless free energy functional representing 
the coupling between the two layers, given by 
\begin{equation}
\ff_{\rm gh} =  a_2 \int d\vec{r}\
(\Delta h-\Delta h^{0})^2 
+ \int d\vec{r}\left(V_{\rm N} n_{\rm N}
+V_{\rm B} n_{\rm B}\right) n_{\rm g},
\label{eq:Fgh}
\end{equation}
where $\Delta h = h_{\rm g}-h_{\rm h}$ and 
\be
\Delta h^0 = \Delta[1+ n_{\rm g}(\alpha_{\rm gN}  n_{\rm N}
+\alpha_{\rm gB} n_{\rm B})].
\label{eq:Dh}
\ee
This form is similar to that reported by Elder \etal \cite{Elder21}, but 
for computational efficiency, the coupling is between $n_{\rm N}$, $n_{\rm B}$ and $n_{\rm g}$ and 
not between the differences from the average densities (i.e., $n_{\rm N}-\bar{n}_{\rm N}$ etc.) 
which were simpler for 
analytic calculations. The parameters entering Eqs.~(\ref{eq:Fgh}) and (\ref{eq:Dh}) 
(i.e., $a_2$, $\Delta$, $\alpha_{\rm gN}$, $\alpha_{\rm gB}$, $V_{\rm N}$, and $V_{\rm B}$)
will be discussed in more detail in the next section.
Finally to allow for out-of-plane deformations the Laplacian entering 
Eqs.~(\ref{eq:fgraph}) and (\ref{eq:fhBN}) becomes
\be
\Lap \rightarrow 
(1-h_x^2)\partial_{xx} +(1-h_y^2)\partial_{yy} -2h_xh_y\partial_{xy},
\label{eq:Laph}
\ee
where $h_i\equiv \partial_i h$ and the $h$ field entering 
Eq.~(\ref{eq:Laph}) is $h_{\rm g}$ in Eq.~(\ref{eq:fgraph}) and 
$h_{\rm h}$ in Eq.~(\ref{eq:fhBN}).  

As discussed in Refs.~\cite{Pete16} and \cite{Taha17},
Eqs.~(\ref{eq:fgraph}) and (\ref{eq:fhBN}) are in essence the original  
PFC model (with some couplings between two components in the hBN case) that are minimized by 
a periodic structure due to the term $(\Lap+q_{\rm X}^2)$,  
where X is g, B or N and $q_{\rm X}$ uniquely determines the lattice periodicity, for a given set of parameters.
The other polynomial terms in the free energy functional are essentially a standard Landau 
expansion that gives rise to two potential wells of differing height, which breaks 
the up-down symmetry and leads to 2D triangular 
patterns as opposed to one-dimensional stripe patterns. 
The corresponding coefficients can be connected to the Fourier components of
the expansion of direct correlation functions in classical DFT \cite{Elder07,Taha17,Taha19}.
For convenience $q_{\rm g}$ was set to unity and as 
discussed in the next section, $q_{\rm B}$ and $q_{\rm N}$ were chosen to give the 
correct graphene/hBN lattice constant ratio.  Details of the choice of the 
parameters entering Eqs.~(\ref{eq:fgraph}) and (\ref{eq:fhBN}) are 
given in Refs.~\cite{Pete16} and \cite{Taha17} respectively.  Details of 
the graphene-hBN layer coupling [i.e., Eq.~(\ref{eq:Fgh})] are discussed in Ref.~\cite{Elder21} as 
well as in the following section. 

The dynamics of the fields are conserved for the densities and 
nonconserved for the heights, i.e., 
\be
\pt{n_\alpha}{t}= \Lap \frac{\delta \ff}{\delta n_\alpha},
\label{eq:ndt}
\ee
and
\be
\pt{h_\alpha}{t} = -\Gamma \frac{\delta \ff}{\delta h_\alpha},
\ee
where the subscript $\alpha = {\rm g, h}$ corresponds to either graphene (g)
or hBN (h).  Since the focus of this work is on equilibrium states, 
$\Gamma$ was chosen as large as possible to get to equilibrium.  
Typically, $\Gamma \approx 10-50$, although in some cases after 
initial relaxation it was possible to increase $\Gamma$ up to $10000$.
In all the following calculations periodic boundary conditions are used.

\section{Equilibrium}
\label{sec:equi}

For simplicity the wave numbers $q_{\rm N}$ and 
$q_{\rm B}$ will be set identical, {\it i.e.}, 
$q_{\rm N}=q_{\rm B}=q_{\rm h}$ in the hBN layer.
In the lowest-order Fourier expansion of the density fields $q_{\rm h}$ is given by 
\be
q_{\rm h} = a_{\rm g}/a_{\rm h},
\ee
where $a_{\rm g}$ and $a_{\rm h}$ are the lattice constants of the 
graphene and hBN layers respectively.  However, as will be
discussed below, $q_{\rm h}$ must be numerically fitted
since higher-order Fourier modes will play a non-negligible role in 
determining the lattice constants.
It is useful to rewrite the dimensionless bilayer free energy functional $\ff$ as 
\be
\frac{\ff}{c_{\rm g}} = \ff_{\rm g} + \ff_{\rm gh}+ \frac{c_{\rm h}}{c_{\rm g}}\ff_{\rm h},
\ee
to incorporate the ratio $c_{\rm h}/c_{\rm g}$ in numerical simulations.
It was found that if $q_{\rm h}$ were set to 
\be
q_{\rm h} =1.011 a_{\rm g}/a_{\rm h},
\ee
the dimensionless lattice constants obtained numerically became
\begin{equation}
a_{\rm h} =  7.4191, \qquad
a_{\rm g} = 7.2721.
\end{equation}
This then gives $a_{\rm g}/a_{\rm h}=0.9802$, which is close to the ratio 
$2.46/2.51=0.9801$ between graphene (of lattice constant $2.46$ \AA) 
and hBN ($2.51$\AA). 

To ascertain the nature of the equilibrium state, a bilayer was 
constructed with $49 \times 49$ unit cells of hBN and $50 \times 50$ 
of graphene.  This turned out to be an unstable initial condition which 
spontaneously relaxed to the graphene layer becoming commensurate 
($49 \times 49$ unit cells) with the hBN layer. 
Next the free energy density of a commensurate
g/hBN bilayer system was examined as a function of the lattice constant 
$a_{\rm b}^{\rm x}$ of the bilayer.  It was found that the lowest energy state 
occurred when 
\be
a_{\rm b}^{\rm x}=7.3548,
\ee
or $2.488$ {\AA} in dimensional units and is slightly closer to the hBN lattice 
constant, i.e., $a_{\rm b}^{\rm x}-a_{\rm g}=0.083$ and $a_{\rm h}-a_{\rm b}^{\rm x}=0.064$.  

\begin{figure}[htb]
\vskip 0pt
\includegraphics[width=0.46\textwidth]{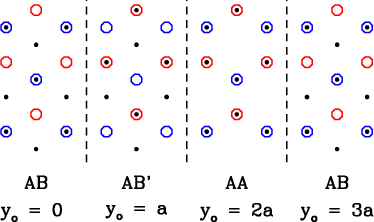} 
\caption{Stacking for the graphene/hBN bilayer. The black, blue and red points 
correspond to density maxima in $n_{\rm g}$, $n_{\rm N}$ and $n_{\rm B}$ respectively.  $y_0$ is 
the shift of the graphene lattice with respect to the hBN lattice and $a$ is the 
nearest-neighbour distance.}
\label{fig:ghBNt}
\end{figure}

\begin{table}[htb]
\begin{tabular}{|c|c|c|}
\hline
Model parameter &  One-mode approximation  & Adjusted value \\
\hline
$V_{\rm N}$ & $2.06 \times 10^{-4}$ & $2.25 \times 10^{-4}$ \\
\hline
$V_{\rm B}$ & $2.64\times 10^{-5}$ & $5.20\times 10^{-5}$ \\
\hline
$\Delta$ & $10.31$ & $10.32 $\\
\hline
$\alpha_{\rm gN}$ & $0.195$ & $0.21$ \\
\hline
$\alpha_{\rm gB}$ & $0.037$ & $0.07$  \\
\hline
$a_2$ & $7.31 \times 10^{-5}$ & $7.31 \times 10^{-5}$  \\
\hline
\end{tabular}
\caption{Summary of model parameters for the graphene/hBN bilayer PFC model.} 
\label{tab:param}
\end{table}

\begin{figure}[htb]
\vskip 0pt
\includegraphics[width=0.46\textwidth]{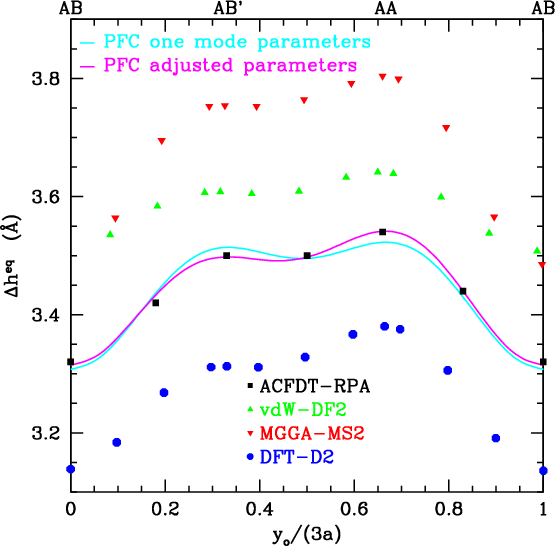}
\caption{Stacking height predictions.  
The DFT calculations (ACFDT-RPA, vdW-DF2, MGGA-MS2, and DFT-D2) 
are from Zhou \etal \cite{Zhou15}.}
\label{fig:stackh}
\end{figure}

\begin{figure}[htb]
\vskip 0pt
\includegraphics[width=0.46\textwidth]{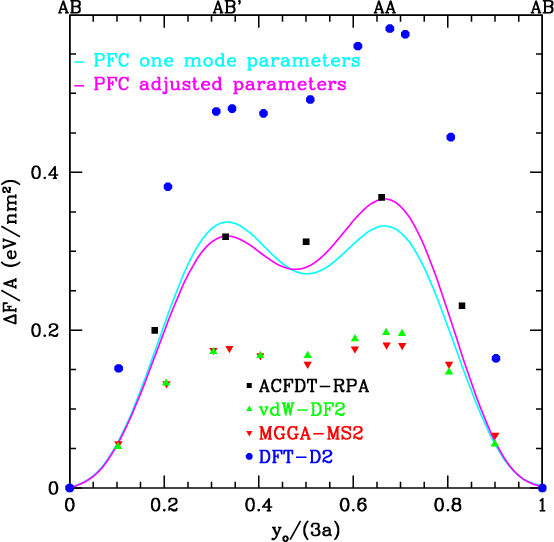}
\caption{Stacking energy density predictions.  
The DFT calculations (ACFDT-RPA, vdW-DF2, MGGA-MS2, and DFT-D2) 
are from Zhou \etal \cite{Zhou15}.}
\label{fig:stackf}
\end{figure}

The free energy density difference $\Delta F/A$ (with area $A$) 
and the relative height $\Delta h^{\rm eq}$ were examined for the bilayer as a 
function of stacking position, where $\Delta F$ is the difference 
with respect to the AB stacking. The stacking is illustrated in 
Fig.~\ref{fig:ghBNt}, for densities $n_{\rm g}(x,y+y_0)$ and $n_{\rm B(N)}(x,y)$ 
such that when $y_0=0$ an AB stacking (the lowest energy state) 
occurs.  The parameters, as listed in Table \ref{tab:param}, were initially 
chosen using the analytic one-mode approximation (which includes only 
the lowest-order Fourier coefficients needed to 
reconstruct the graphene and hBN crystalline lattices) as described by 
Elder \etal \cite{Elder21}. They were obtained by fitting to 
the DFT calculations of Zhou \etal \cite{Zhou15}
which considered four different DFT approaches and determined the one 
with the acronym ACFDT-RPA giving the best predictions for bulk properties;
as such these data were used to fit the current PFC model. 
Their predictions for $\Delta F/A$ and equilibrium $\Delta h^{\rm eq}$
are shown in Figs.~\ref{fig:stackh} and \ref{fig:stackf} respectively. 
Numerical simulations of the PFC model were conducted (which naturally 
include all Fourier coefficients) to minimize the free energy for an AB 
stacking.  This configuration was then used to determine the energy 
of other stackings as described by Elder \etal \cite{Elder21}. 
As with the DFT calculations these were done on a single 
unit cell which does not allow for out-of-plane deformations.  
The outcomes of these studies show that results from the one-mode parameters are close to 
the DFT calculations but are slightly different with small deviations
for the height and free energy density difference between the AB$'$ and 
AA stackings, as seen in Figs.~\ref{fig:stackh} and \ref{fig:stackf}.
One particular feature is that in the one-mode predictions 
the magnitude of $\Delta h^{\rm eq}$ and $\Delta F/A$ are very similar for 
the AB$'$ and AA stackings which however are clearly different 
in all the DFT calculations. For this reason the parameters were 
adjusted to obtain a better fit as shown in both Figs.~\ref{fig:stackh} 
and \ref{fig:stackf}. A summary of the corresponding dimensionless 
parameters obtained are given in Table \ref{tab:param}. These adjusted 
parameters are used in all the subsequent simulations that follow.

\section{Inversion Domain Boundaries in hBN and the graphene/hBN bilayer}
\label{sec:IDB}

\begin{figure}[htb]
\vskip 0pt
\includegraphics[width=0.35\textwidth]{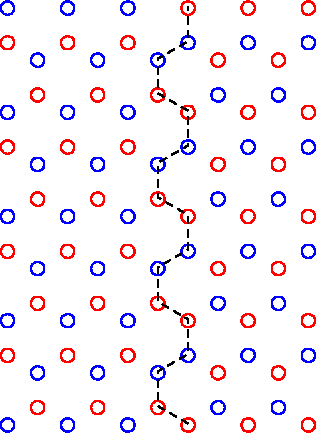}
\caption{Illustration of an unstable inversion domain boundary.  The blue 
and red dots indicate N and B atoms respectively and the dashed 
black line highlights the inversion boundary.}
\label{fig:inver}
\end{figure}

\begin{figure}[htb]
\centerline{
\includegraphics[width=0.5\textwidth]{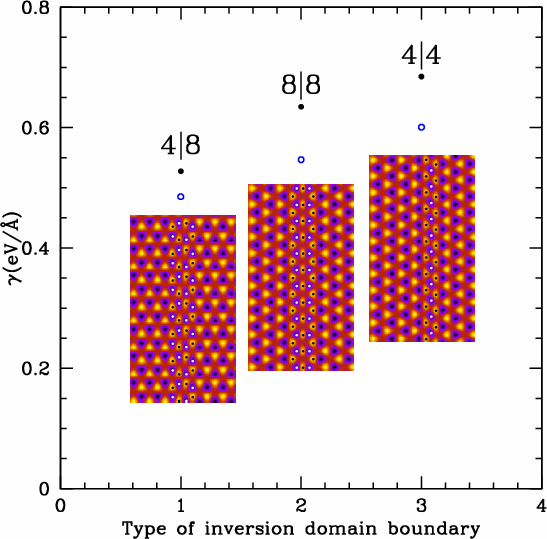}}
\caption{Grain boundary energy $\gamma$ of three lowest-energy inversion 
domain boundaries in hBN monolayers. The 
solid black and open blue dots above the snapshots correspond to rigid and flexible 
sheets respectively, where the energy was minimized with respect to the dimensions 
of the simulation box. Note that only one unit cell was included in the $y$ direction.
In the configurations, blue and yellow maxima 
correspond to the positions of the N and B atoms, respectively, 
and black and white dots have been placed on lattice locations at 
the inversion boundaries.}
\label{fig:inversion}
\end{figure}

In a prior publication \cite{Taha17} an examination of inversion 
boundaries in hBN was done using a rigid model, i.e., using the free energy 
functional in  Eq. (\ref{eq:fhBN}) with $h_{\rm h}=0$. An inversion boundary forms when 
the atomic ordering switches from BNBNBN to NBNBNB as illustrated 
in Fig.~\ref{fig:inver}.  As drawn in the figure the boundary 
contains many homoelemental nearest neighbours in the middle portion, 
which would be very unfavorable energetically.  Instead the system prefers 
to form defected structures with unit rings that contain
more or less than six atoms, but avoid having homoelemental 
B$-$B or N$-$N neighbouring. In particular, the grain boundary energy per unit 
length ($\gamma$) of inversion boundaries that contain $4|8$, $8|8$,
and $4|4$ defect structures will be studied, where $i|j$ corresponds 
to neighbouring defect pairs containing $i$- and $j$-membered atomic 
rings.  In prior work \cite{Taha17} it was found that the $4|8$ boundary 
naturally emerges 
when the boundary is along the armchair (AC) direction, while the 
$8|8$ results from the zigzag (ZZ) orientation.  There was also a
$4|4$ boundary along a ZZ interface that was slightly shifted;
hence strictly speaking the $4|4$ is not 
an inversion boundary due to the shift (see Fig.~\ref{fig:inversion}). 

In this section inversion boundaries will be examined for a flexible 
hBN monolayer  and a hBN/graphene bilayer.  These studies will illustrate 
the impact of allowing out-of-plane deformations on a single layer as well 
as the influence of the hBN inversion boundary on the graphene layer in 
a bilayer  system.

Simulations were first conducted to reproduce the results of Taha \etal \cite{Taha17}. 
A fully periodic box of size 
$L_x \Delta x \times L_y \Delta y$, where $L_x$ and $L_y$ 
are integers, was used.
In these simulations a box of grid size 
$3200 \Delta x \times 24 \Delta y$
for the AC configuration and 
$3200 \Delta x \times 14 \Delta y$
for the ZZ configuration were used.  This corresponds to boxes of 
size 
$563$ \AA\ $ \times\ 4.22$ \AA
for the AC and 
$563$ \AA\ $ \times\ 2.46$ \AA
for ZZ, and corresponds to a
single unit cell in the $x$ direction.  It should be noted that 
conserved dynamics (i.e., Eq.~(\ref{eq:ndt})) were employed, 
which do not fix the local density while ensuring a constant 
average density in the whole system. Taha \etal \cite{Taha17} typically 
found grain boundary energies saturate for system sizes of $500$ \AA\ and larger.
For the rigid case (without out-of-plane deformations) the initial 
condition was such that half the simulation box was of configuration NBNB 
while the other half was BNBN with a uniform density band of width 
$20 \Delta x$ placed at the boundaries. Simulations were run until 
the system energy was minimized.  Next, $\Delta x$ and $\Delta y$ were 
varied to find the minimum energy state (since it is not possible to know the 
desired width of the domain walls).  In the simulations here (flexible 
hBN monolayer and hBN/graphene bilayer) the same procedure was followed.
Fixing $h=0$ reproduced the results  of Taha \etal \cite{Taha17} for the 2D 
rigid planar systems.

\begin{figure}[htb]
\vskip 0pt
\includegraphics[width=0.48\textwidth]{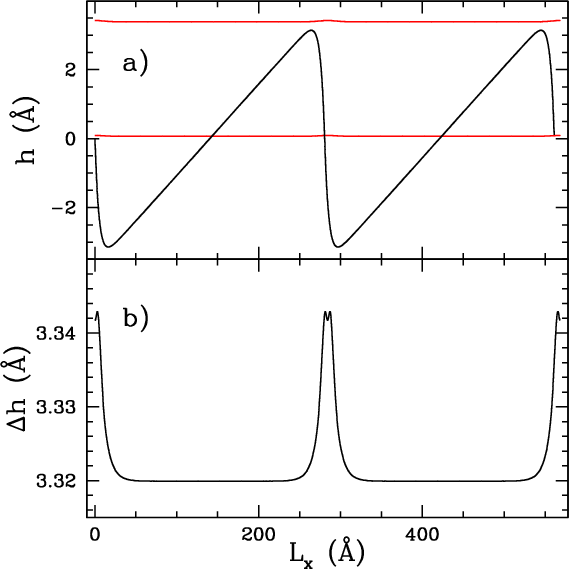}
\caption{(a) Comparison of height variation across two 
$4|8$ inversion boundaries in a single monolayer of hBN (black curve) 
and a g/hBN bilayer (red lines), where the 
upper (lower) red line is for graphene (hBN) in the bilayer.
(b) Height difference between the graphene and hBN layers in the g/hBN 
bilayer ({\it i.e.}, the difference between the red lines in (a)).
}
\label{fig:heights}
\end{figure}

Following this simulations were conducted for a flexible sheet allowing out-of-plane
deformations (i.e., containing variations in $h$). A first test was conducted on a system 
of size $563$ \AA\ $\times\ 169$ \AA, with the initial condition
set up by reproducing $40$ lattice constants for an AC boundary
along the $x$ direction.  The initial height was set to be 
a uniform random number in the range $-1/8 < h < 1/8$.
The simulation results showed that the height developed into a 
one-dimensional pattern perpendicular to the domain wall (see Fig.~\ref{fig:heights}(a)).  
This indicates that one unit cell along the parallel direction of this 
pattern was sufficient.  Allowing for out-of-plane deformations lowered 
the domain wall energy, as shown in Fig.~\ref{fig:inversion}. 
Despite a modest bending of the sheet (of the order of one atomic spacing, 
as seen in, e.g., Fig.~\ref{fig:heights}(a)), a considerable decrease of 
system energy was observed (by 8.0\%, 13.9\%, and 9.4\% respectively for 
$4|8$, $8|8$, and $4|4$ boundaries). 

\begin{figure}[htb]
\vskip 0pt
\includegraphics[width=0.48\textwidth]{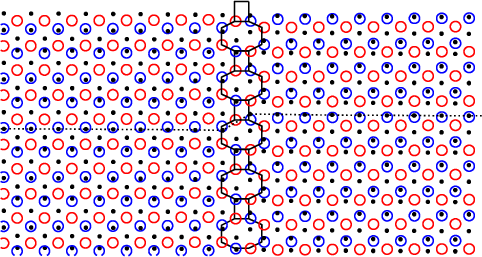}
\caption{Structure of inversion boundary in a g/hBN bilayer obtained 
from numerical simulations.  The color/point scheme is the same as that in 
Fig.~\ref{fig:ghBNt}.  The dashed line is guide to the eye to highlight the 
\gbound in graphene lattice across the inversion boundary.
Note that only one unit cell is simulated in the $y$ 
direction.}
\label{fig:Oacednjb}
\end{figure}

A further simulation was conducted to understand the influence of such boundaries in g/hBN bilayers. The initial condition for the hBN 
layer was the same as that described for the monolayer case, except the lattice 
constant used ($2.488$ \AA) was the one that minimizes the g/hBN bilayer 
as discussed in Sec.~\ref{sec:equi}. The graphene layer 
was initialized to be in an AB stacking configuration on both sides of 
the grain boundary. Near the boundaries the graphene density was 
set to be uniform in a width of $20 \Delta x$ 
so that a boundary would naturally form.  This leads to 
an inversion boundary in the hBN layer with $4|8$ defect 
pairs and a \gbound in the graphene layer, in which the 
graphene slides perpendicular to boundary, as 
illustrated in Fig.~\ref{fig:Oacednjb}.  As can be seen in this figure 
the graphene layer remains in the AB stacking but the atomic sites 
need to shift across the boundary, leading to in-plane local distortions 
and strain in the graphene layer.  Initially the heights in both layers 
resemble that of the monolayer case; however as time evolves 
larger height gradients appeared at the boundaries, 
leading to numerical instabilities.  In order to eliminate the 
instability the grid spacings in both the $x$ and the $y$ directions
were reduced by a factor of two. With this change during the time evolution
the heights spontaneously transformed into a much 
smoother profile eventually, as shown in Figs.~\ref{fig:heights}(a) 
and \ref{fig:heights}(b).

The free energy difference  per unit length was measured 
to be $\gamma=0.471$ eV/{\AA} for the inversion boundary 
in the g/hBN bilayer system.  While it is 
somewhat interesting that this value is comparable to the 
single-layer result of hBN,
it is important to note that in the single-layer system the inversion 
boundary energy was compared to an unstrained, flat equilibrium 
hBN layer (with lattice constant $a_{\rm h}=2.51$ \AA), while for the 
bilayer it was compared to a configuration with lattice 
constant $2.488$ {\AA} that minimizes the 
bilayer, where both the graphene and hBN layers are strained in the 
lowest energy state.  This significantly restricts out-of-plane deformations 
since the hBN layer is under compression and graphene under tension.
In addition, the bilayer is a three-dimensional 
system.  The three-dimensional inversion boundary energy, $\gamma_{3d}$, is 
$0.471/3.32$ eV/\AA$^2$ = 0.142 eV/\AA$^2$, where $3.32$ \AA\ is the 
vertical spacing between the layers.

While these simulations give insights of the influence of a 
defect (i.e., inversion boundary in hBN) in the coupled layers, it should 
be noted that the quantitative results will depend on the specific setup.  
In the above simulations an inversion boundary was formed in hBN 
at the equilibrium lattice constant (2.488 \AA) of the g/hBN bilayer system.   
This would be quantitatively different from growing graphene on an 
already formed hBN inversion boundary that was at the lattice constant  (2.51 \AA) 
of a single hBN layer. In addition, the procedure in the 
simulations conducted here was to adjust the width and length of the system 
to minimize the free energy.
Presumably changes in the setup and procedure could lead to quantitatively 
different results, although they are unlikely to change the qualitative 
structures in the two layers, such as the \gbound and its resulting strain in the 
graphene layer.

\begin{figure}[htb]
\vskip 0pt
\includegraphics[width=0.48\textwidth]{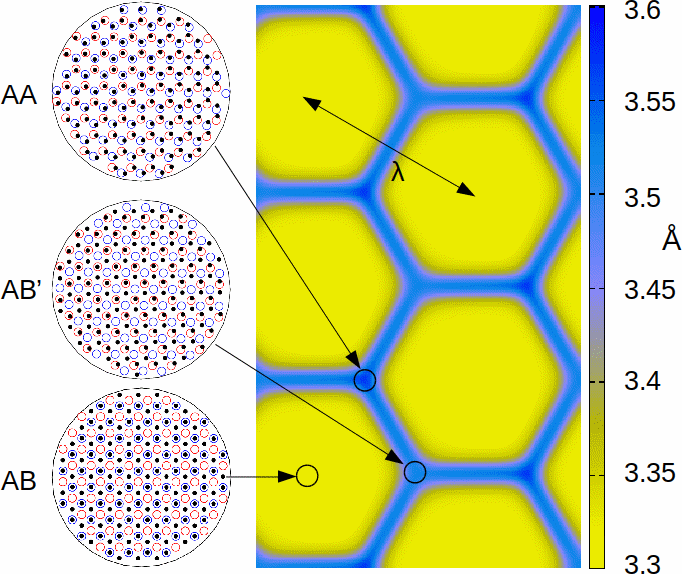}
\caption{Sample configuration for misorientation $2\theta = 0.66^\circ$ 
(for $k=150)$ and pattern wavelength $\lambda=21.4$ nm. The color scheme 
on the right corresponds to the height difference between the layers. 
 The size of the configuration is $37.0$ nm $\times$ $64.1$ nm.
On the left the atomic configurations are shown 
for the circled regions. The color scheme is the same as that of Fig. 1.}
\label{fig:NNN150H}
\end{figure}

\section{Twisted Bilayers}
\label{sec:bilayers}

To study a twisted bilayer of graphene and hBN numerically,  
one layer was rotated by an angle $\theta$ and the other by $-\theta$, 
for a total misorientation of $2\theta$, with both lattice constants 
set to those identified in Sec.~\ref{sec:equi}, i.e., 2.488\AA. This gives rise to the
appearance of Moir\'e patterns and some interesting behaviour as 
observed in many previous works \cite{Elder21,Zhou15,Elder16,Elder17}.  
As discussed in Ref.~\cite{Elder21} only certain rotation angles 
and box sizes can be used in a periodic simulation box.  More 
precisely, given an integer $k$ in the zigzag orientation with 
rotation angle
\be
\tan \theta = \frac{\sqrt{3}}{2(k+1/2)},
\ee
the system size $(L_x,L_y)$ must be 
\be
L_x &=& a_x \sqrt{3/4+(k+1/2)^2}; \ L_y = \sqrt{3}L_x,
\ee
where $a_x = 4\pi/(\sqrt{3}\,q_{\rm min})$.  A typical configuration 
at a small angle is presented in Fig.~\ref{fig:NNN150H}.  
The wavelength, $\lambda$, of the Moir\'e pattern shown in this 
figure is given by $\lambda=L_x/\sqrt{3}$. The figure 
indicates that the height difference is the smallest 
in the AB stacking regions and the largest at the AA junctions, 
followed by the AB$'$ junctions, consistent 
with Fig.~\ref{fig:stackh}.  This makes the pattern slightly 
non-symmetric or tilted.  

\begin{figure}[htb]
\vskip 0pt
\includegraphics[width=0.48\textwidth]{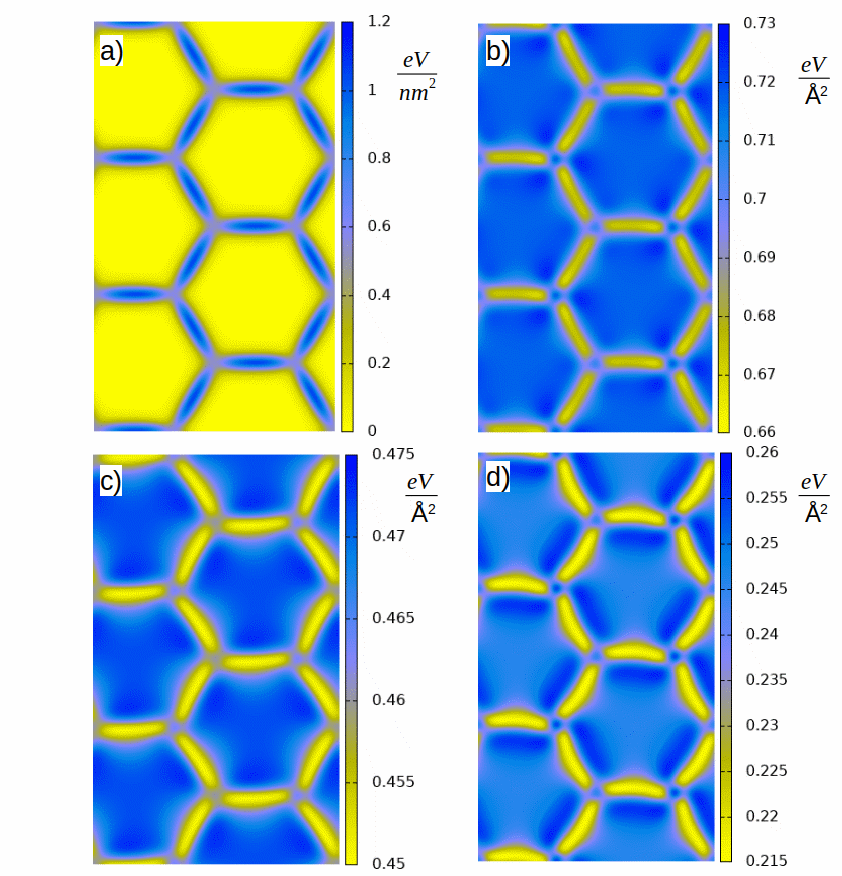}
\caption{Sample configurations at misorientation $2\theta=0.66^\circ$, 
for (a) smoothed free energy density difference and the smoothed
volumetric stress in (b) the combined bilayer, (c) graphene and 
(d) hBN layers respectively.  The scale is in units of eV/nm$^2$ 
in (a) and eV/\AA$^2$ in (b), (c), and (d).  The size of the system 
is the same as that of Fig. \ref{fig:NNN150H}.}
\label{fig:NNNtwst150_Vall}
\end{figure}

Sample configurations are shown in Fig. \ref{fig:NNNtwst150_Vall} 
for the corresponding free energy density and the volumetric stress, $\sigma_{\rm V}=\sigma_{xx}+\sigma_{yy}$, 
of the whole system and individual layers.  These quantities vary 
on the length scale of the atomic spacing, which makes it difficult to 
observe the overall pattern.  For this reason in the visualization of patterns they were 
smoothed via the multiplication of $e^{-\alpha_0 k^2}$ in Fourier space,
where $k$ is the wave number, and then an inverse Fourier transform.  
A value of $\alpha_0=14$ was found to mostly eliminate the 
small scale oscillations while not washing out the large 
scale Moir\'e patterns, and was used for the pattern visualization of all angles.
Figure \ref{fig:NNNtwst150_Vall} shows that  
the triple junctions in the pattern are slightly twisted, particularly 
evident in the $\sigma_{\rm V}$ spatial distribution of individual layers 
(see panels (c) and (d)), from which it is also interesting to note
that the junctions in two layers twist in opposite directions.  
Similar twisted junctions have been observed in many other 
strained-layer Moir\'e patterns 
\cite{Elder16,tumino11,wu11,ling06,dai16,pushpa03,corso10}. 
This twisting occurs to move the junctions to lower the junction 
energy, even though this slightly increases the length 
of the domain walls connecting the junctions.

\begin{figure}[htb]
\vskip 0pt
\includegraphics[width=0.48\textwidth]{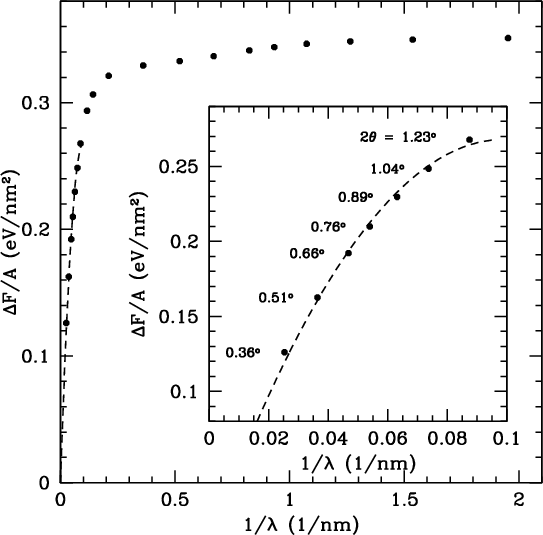}
\caption{Free energy density difference as a function of $1/\lambda$. 
The points correspond to simulation data and the dashed line corresponds 
to a second-order polynomial fit to $\Delta F/A$ in terms of $1/\lambda$ 
according to Eq.~(\ref{eq:F_lambda}) for $1/\lambda < 0.10~{\rm nm}^{-1}$.}
\label{fig:NTwistF}
\end{figure}

\begin{figure}[htb]
\vskip 0pt
\includegraphics[width=0.47\textwidth]{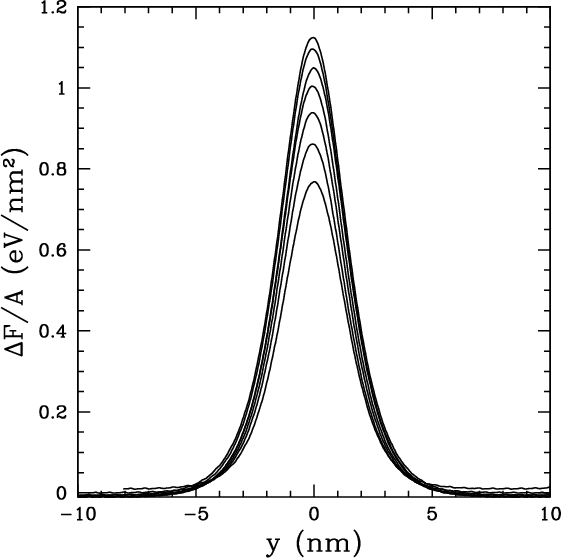}
\caption{Free energy density difference across a domain wall. 
The lines from top to bottom correspond 
to angles $2\theta=0.36^\circ, 0.51^\circ, 0.66^\circ, 0.76^\circ, 0.89^\circ,1.04^\circ$, and $1.23^\circ$.}
\label{fig:Wallepx}
\end{figure}

For very small angles it is possible to 
postulate that the system free energy would scale as 
$\ff =  2\jeng + (3\lambda/\sqrt{3}) \gamma + A f_c$, 
where $\jeng$ is the free energy of the junction, $\gamma$ is the 
energy per unit length of the domain wall, $f_c$ is the 
free energy density of the commensurate regions, and 
$A=(\sqrt{3}/2)\lambda^2$ is the area of a hexagon in the pattern.  
The factor of $2$ in front of $\jeng$ is due to the fact that 
each junction contributes $\jeng/3$ to each hexagon and there 
are six junctions per hexagon.  The $3\lambda/\sqrt{3}$ factor
in front of $\gamma$ arises from the fact that each domain wall 
length is $\lambda/\sqrt{3}$ and that there are six domain walls 
per hexagon with each wall contributing to two hexagons.
This then implies that the free energy density difference scales as 
\be
\frac{\Delta \ff}{A} =  \frac{4}{\sqrt{3}}\frac{\jeng}{\lambda^2}
+\frac{2\gamma}{\lambda}. \label{eq:F_lambda}
\ee
The total free energy density of the bilayer is shown as a function 
of the inverse periodicity ($1/\lambda$) of the pattern in 
Fig.~\ref{fig:NTwistF}, which includes a fit to the form 
$\Delta \ff/A =  \alpha/\lambda+\beta/\lambda^2$ for the small angle 
data.  This gives a prediction for the junction energy 
$\jeng = -11.9$ eV and domain wall energy density $\gamma = 2.71$ eV/nm.  
These results should be taken with a grain of salt as they assume 
$\lambda$ is much larger than the size of the defects (domain walls 
and junctions) and that the specific form of defects does not change 
with system size.  However, it is clear that there are in fact changes 
in these defects as shown in Fig. \ref{fig:Wallepx}, where the free
energy density across a domain wall can be seen to increase with 
larger system size.

\begin{figure}[htb]
\vskip 0pt
\includegraphics[width=0.46\textwidth]{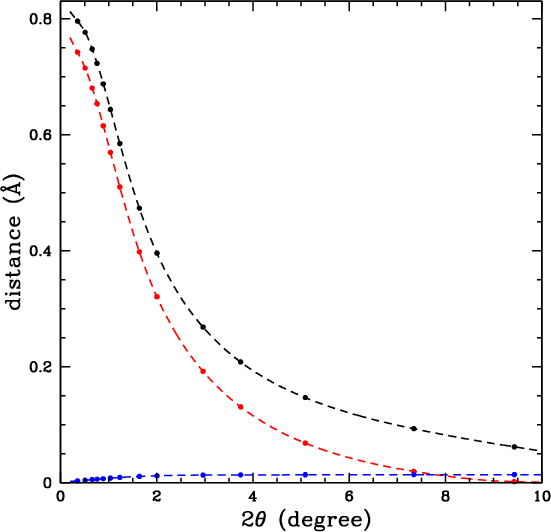}
\caption{Buckling of the hBN (black line) and graphene (red line) layers
as a function of misorientation.  In addition, the difference of the 
average distance between layers with respect to the equilibrium AB-stacking 
distance (i.e., $\langle h_{\rm h}-h_{\rm g} \rangle -\Delta h_{\rm AB}$) is shown in blue.  
The lines are guides to the eye.}
\label{fig:heightp}
\end{figure}

The change in free energy density is also accompanied by a change in the 
buckling of the individual layers.  In Fig.~\ref{fig:heightp} the buckling 
width $\sqrt{\langle (h_\alpha-\bar{h}_\alpha)^2 \rangle}$ (where $\alpha$ refers to the 
graphene (g) or hBN (h) layer) is depicted.  The buckling reaches a maximum as 
$\theta \rightarrow 0$ and becomes very small for large angles as has 
been observed in other studies \cite{Kumar15}.   Also shown in this 
figure is the average distance between the layers minus the equilibrium 
AB-stacking distance. As can be seen this distance is much smaller than 
the buckling of the individual layers, indicating that the sheets buckle 
in sync with each other as has been observed in graphene/graphene bilayers 
\cite{Zhou15,dai16,Elder21}.

\begin{figure}[htb]
\vskip 0pt
\includegraphics[width=0.48\textwidth]{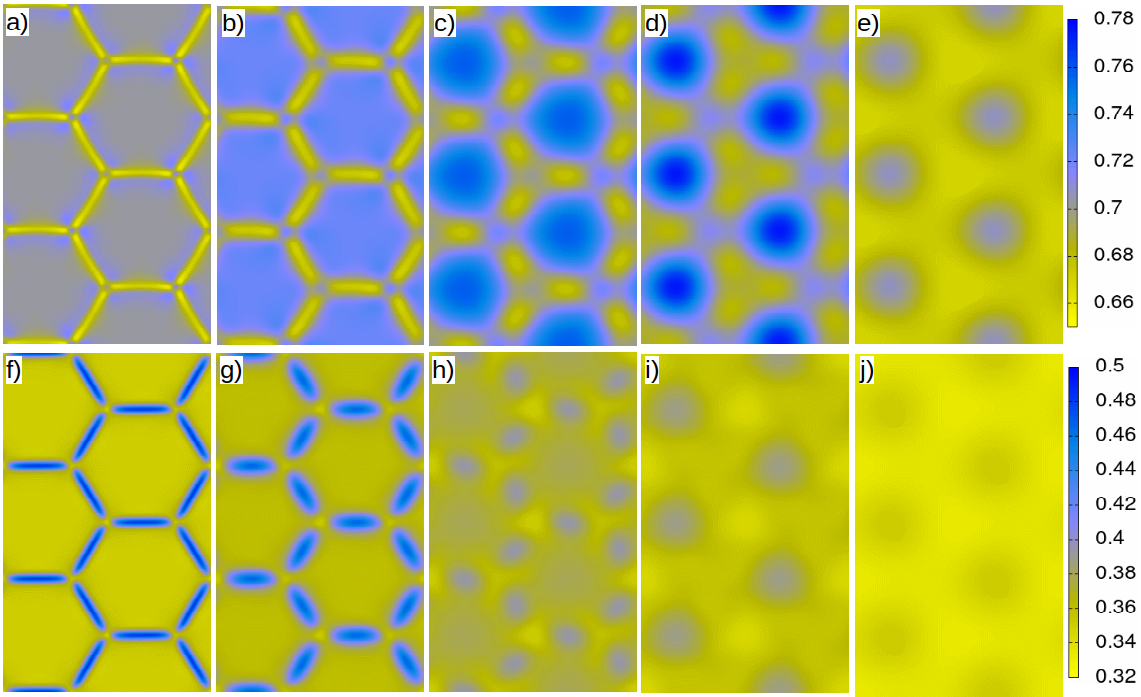}
\caption{Smoothed volumetric [(a)--(e)] and von Mises [(f)--(j)] stresses 
for misorientation angles and pattern wavelengths $(2\theta,\lambda)=
(0.36^\circ,39.6\,{\rm nm})$,  
$(0.76^\circ,18.5\,{\rm nm})$, 
$(1.64^\circ,8.5\,{\rm nm})$, 
$(2.96^\circ,4.8\,{\rm nm})$, and 
$(9.43^\circ,1.5\,{\rm nm})$ for (a) to (e) or (f) to (j) respectively. 
The color scale is in units of eV/nm$^2$. 
In each instance the system size is 
$\sqrt{3}\lambda \times 3 \lambda$.}
\label{fig:Strains}
\end{figure}

\begin{figure}[htb]
\vskip 0pt
\includegraphics[width=0.46\textwidth]{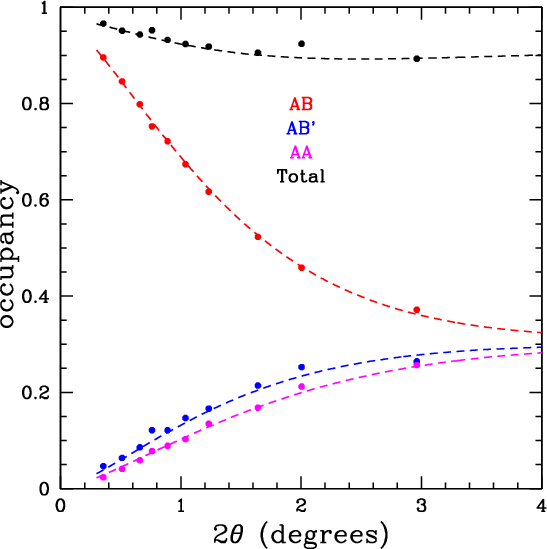}
\caption{Occupancy of AA, AB$'$, and AB sites as a function of misorientation. 
The black points correspond to the sum of the AB, AB$'$, and AA occupancies.
The lines are guides to the eye.}
\label{fig:phases}
\end{figure}

\begin{figure}[htb]
\vskip 0pt
\includegraphics[width=0.46\textwidth]{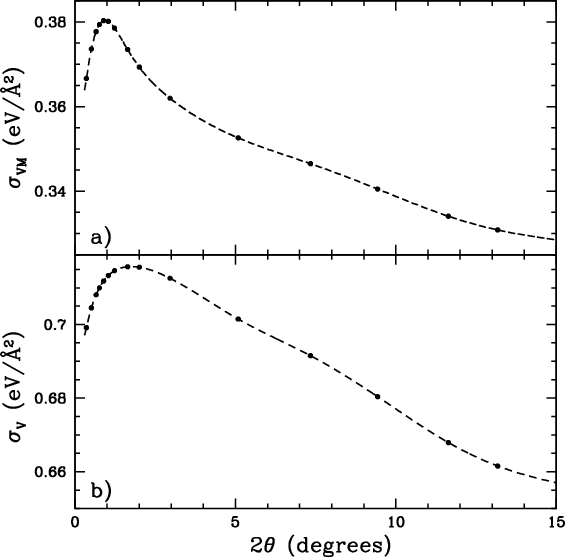}
\caption{Average von Mises and volumetric stresses as a function of misorientation 
in (a) and (b) respectively.}
\label{fig:SigVVm}
\end{figure}

More interesting than the buckling is the change in stress as a function of 
misorientation.  The picture of well-defined domain walls and junctions 
breaks down for large misorientations when the domain walls become comparable 
with the size of the Moir\'e pattern. This can be well captured by the 
volumetric ($\sigma_{\rm V}$) and von Mises \cite{Dantzig09}
($\sigma_{\rm VM})$ stresses, where the latter is given by 
$\sigma_{\rm VM} = \sqrt{\sigma_{xx}^2+\sigma_{yy}^2-\sigma_{xx}\sigma_{yy}+3\sigma_{xy}^2}$
in 2D after neglecting the $z$-direction stress components.
The corresponding results are shown in Fig.~\ref{fig:Strains}.
The patterns exhibit a transition 
from those with well-defined triple junctions and domain walls to smeared-out 
patterns around $\theta \approx 0.5^\circ-0.8^\circ$ (with $\lambda=15-25$ nm). 
As can be seen in Fig. \ref{fig:Wallepx} the size of the domain wall is on the 
order of $5-10$ nm; thus the transition occurs roughly when the domain walls 
begin to overlap.  In this case there are no well-defined AB, AB$'$ and AA 
regions as indicated in Fig. \ref{fig:phases}, which shows the occupancy 
of these states as a function of $\theta$.  It should be noted that it 
was not always possible to determine the state of a given unit cell; 
i.e., as shown in this figure the sum of the occupancy does not add to one.
Clearly for small angles the AB states dominate since they are the lowest 
energy phases, followed by the next lowest energy state AB$'$ and finally 
by the AA state as expected.  This also implies that for small angles 
the AA junctions are slightly smaller than the AB$'$ junctions.

The transition from small to large angle patterns can be identified from the 
average von Mises and volumetric stresses as given in Fig.~\ref{fig:SigVVm}.  
Both become larger with the increase of misorientation angle, then reach
a peak before slowly decreasing.  The peak in $\sigma_{\rm VM}$ occurs at $2\theta=1.03^\circ$
and in $\sigma_{\rm V}$ appears at $2\theta=1.76^\circ$,
consistent with the observation in spatial profiles of stress distribution
in Fig.~\ref{fig:Strains} which show the
transition between two different types of Moir\'e patterns.  
However, it is interesting to note the differences between the patterns. 
In the volumetric case the stress is largest in the commensurate regions and smallest 
at the domain walls.  This is the exact opposite of the von Mises stress, which is 
most apparent at small misorientations, as seen in Fig.~\ref{fig:Strains}. 
The spatial difference between the two stresses is likely 
due to the fact that von Mises stress
incorporates effects of distortion and shearing, and thus would be large at domain walls,
while volumetric (hydrostatic) stress accounts for the effect of volume change but not
distortion or shear, and thus would be small at domain boundaries but large in the domain
bulk subjected to lattice compression or tension.
In addition, in the volumetric case the stress increases in the commensurate 
regions with increasing angle until the transition to the smeared-out state occurs.  
In contrast, the von Mises stress in the commensurate regions does not vary 
as much with the change of misorientation.  It appears that as the angle 
decreases from large values, domain walls emerge, which increases the von Mises 
stress until the walls are fully formed.  When the angle is further reduced, the total 
von Mises stress decreases as the relative area of domain wall compared to 
well-separated commensurate region 
becomes smaller. This results in a maximum of average stress around a transition angle as shown in Fig.~\ref{fig:SigVVm}. This change of von Mises stress distribution indicates 
the change of mechanical property (e.g., yielding) of the bilayer
across the transition of Moir\'e pattern.

\begin{figure}[htb]
\vskip 0pt
\includegraphics[width=0.48\textwidth]{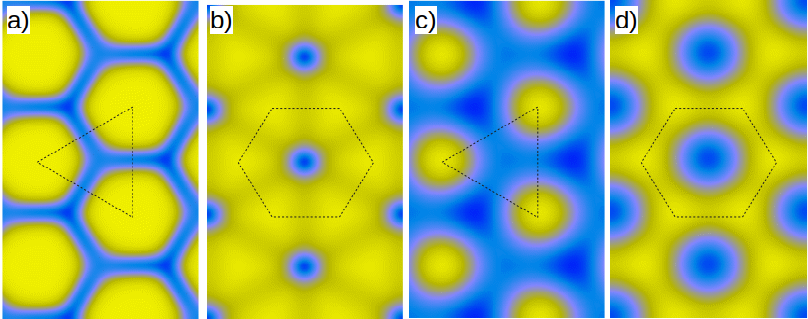}
\caption{Comparison of height differences at misorientations 
$2\theta=1.04^\circ$ in (a) and (b) and $2\theta=3.74^\circ$ in 
(c) and (d).  (a) and (c) correspond to the graphene/hBN bilayer and 
(b) and (d) to the graphene/graphene bilayer.  The dotted black line 
is a guide to the eye to illustrate the triangular and hexagonal 
ordering of the commensurate regions in the graphene/hBN and 
graphene/graphene bilayers respectively.
In (a) and (c) the scale varies 
from 3.25 {\AA} (yellow/light) to 
3.6 {\AA} (blue/dark) and in (b) and (d) 
from 3.25 {\AA} to 3.55 {\AA}.  The system sizes 
are (a) $23.5$ nm $\times$ 40.7 nm,  
(b) $23.2$ nm $\times$ 40.2 nm,  
(c) $6.5$ nm $\times$ 11.3 nm, 
and (d) $6.4$ nm $\times$ 11.2 nm.}
\label{fig:ghBNgg}
\end{figure}

As a final note it is interesting to contrast these results with the 
Moir\'e patterns that appear in graphene/graphene bilayers.  The main 
difference is that in a graphene/graphene bilayer the AB$'$ stacking would 
have an energy identical to the AB stacking, i.e., in such a bilayer no 
AB$'$ junction would exist.  This completely changes the symmetry of the 
domain wall and defect structures from triangular in graphene/graphene 
bilayers to honeycomb shape in g/hBN bilayers.  A comparison of 
the two different systems is shown in Fig.~\ref{fig:ghBNgg} for two 
different misorientations.  The commensurate regions (with lowest height 
differences, appearing yellow/light in the figure) 
of these two types of bilayers have inverse symmetry with respect to each other, 
i.e., a triangular pattern forms in the g/hBN bilayer and honeycomb in 
the graphene/graphene bilayer.

\section{Summary and Conclusions}
\label{sec:summary}

In this work a 2D PFC model incorporating out-of-plane deformations was 
examined for hBN and graphene/hBN bilayers.  In the bilayer case, the model
was parameterized numerically to closely match the ACFCT-RPA DFT calculations 
for stacking energies and height differences between the graphene and hBN
layers obtained by Zhou \etal \cite{Zhou15}, which improves the previous
analytic one-mode calculations of Ref.~\cite{Elder21}.
It was shown that out-of-plane deformations lead to significantly lower 
inversion boundary energies in hBN, on the order of $\approx 8\%-14$\%.  
The boundary in the g/hBN system results in the formation of a \gbound 
with local distortions in the graphene lattice. This interesting defect 
configuration in the g/hBN bilayer gives a domain wall energy of 
$\gamma_{3d}=0.142$ eV/\AA$^2$ as predicted from this PFC calculation.

Numerical simulations were conducted to examine the Moir\'e patterns 
that form when the bilayers are rotated with respect to each other,
showing regions of different types of stacking positions between the layers. 
For small rotations the patterns consisted of well-defined hexagon-shaped
domain walls with triple junctions twisting in opposite directions
in graphene versus hBN layers. Results of the system free energy density,
layer height difference, buckling, and smoothed volumetric and von Mises 
stresses have been obtained for a range of bilayer misorientation angles 
(and Moir\'e pattern wavelengths) that go beyond previous studies. 
An interesting phenomenon observed is the breakdown of well-distinguished 
domain wall structures in the Moir\'e pattern at large enough misorientation
when the domain wall width and the pattern size are of compatible scale,
leading to the transition to a different type of smeared-out Moir\'e pattern 
with overlapping domain boundaries. The corresponding elastic variations of 
these bilayer systems, in terms of volumetric and von Mises stresses, have
been identified, serving as a useful way to characterize the Moir\'e pattern,
transition, and the mechanical property of this type of vertical heterostructures.

\begin{acknowledgments}
K.R.E. acknowledges support from the National Science Foundation 
(NSF) under Grant No. DMR-2006456 and Oakland University Technology Services 
high performance computing facility (Matilda). Z.-F.H. acknowledges
support from NSF under Grant No. DMR-2006446. T.A-N. has been supported in part by the Academy of Finland
through its QTF
Center of Excellence program grant no. 312298. K.R.E. also acknowledges 
useful discussions with M. Greb.
\end{acknowledgments}

\bibliography{graphhBN}
\end{document}